\documentclass[
]{ceurart}

\usepackage{listings}
\usepackage{graphicx}
\usepackage{booktabs}
\usepackage{tabularx}
\usepackage{placeins}
\begin{document}

\copyrightyear{2026}
\copyrightclause{Copyright for this paper by its authors.
  Use permitted under Creative Commons License Attribution 4.0
  International (CC BY 4.0).}

\conference{LASI Spain 26: Learning Analytics Summer Institute Spain 2026, Zamora, May 07--08, 2026}

\title{AISSA: Implementation and Deployment of an AI-based Student Slides Analysis tool for Academic Presentations}


\author[1]{Alvaro Becerra}[%
orcid=0009-0003-7793-2682,
email=alvaro.becerra@uam.es,
] \cormark[1]
\author[1]{Diego Gomez}[%
]
\author[1]{Ruth Cobos}[%
orcid=0000-0002-3411-3009,
email=ruth.cobos@uam.es,
]

\address[1]{GHIA Group, School of Engineering, Universidad Autónoma de Madrid, Spain}
\cortext[1]{Corresponding author.}


\begin{abstract}
Providing timely and actionable feedback on oral presentation slides is challenging in higher education, particularly in large classes where teachers cannot realistically deliver detailed formative feedback before students present. This paper introduces AISSA (AI-based Student Slides Analysis tool), a web-based system that combines large language models (LLMs) and Learning Analytics dashboards to support scalable, rubric-based feedback on presentation slides. AISSA allows students to upload their slide decks prior to an oral presentation and automatically receive quantitative scores and qualitative feedback based on teacher-defined evaluation rubrics. The system analyzes both slide-level features and slide content, generates structured feedback through an LLM (ChatGPT 5.2), and presents the results through interactive dashboards for students and teachers. We tested AISSA on a pilot deployment with 46 undergraduate students in a real academic setting. The results indicate that AISSA is technically reliable, economically feasible, and perceived by students as useful for iterative slide improvement. These findings suggest that combining LLM-based analysis with Learning Analytics dashboards is a promising approach for supporting formative feedback on presentation slides at scale.
\end{abstract}


\maketitle

\section{Introduction}
Feedback is widely recognized as a fundamental pillar of effective teaching and learning. By informing students about the gap between their current performance and intended learning goals, feedback supports reflection, self-regulation, and the improvement of future performance. When designed effectively, it reinforces correct understanding, facilitates error correction, and promotes deeper learning, particularly when it is timely, specific, and aligned with transparent evaluation criteria \cite{hattie2007power, schartel2012giving}. 

Despite its central role in higher education, delivering effective feedback remains a persistent and well-documented challenge \cite{henderson2019challenges}. This difficulty is especially pronounced in the context of oral presentations, where feedback is often limited to a final performance review and therefore arrives too late for students to revise their visual materials beforehand \cite{van2017impact}. Providing detailed, formative, and rubric-aligned feedback on slide design and content prior to the presentation is rarely feasible for teaching staff, particularly in massified courses and under severe time constraints.

To address these limitations, this paper introduces AISSA (AI-based Student Slides Analysis tool), a web-based system that combines large language models (LLMs) and Learning Analytics dashboards to support scalable formative assessment of presentation slides. AISSA allows students to upload their slides before presenting and automatically receive rubric-based quantitative scores together with qualitative, actionable feedback. The system analyzes both slide-level features and slide content itself, using a rubric designed by teachers. The resulting feedback is then presented through interactive dashboards that allow students to inspect scores in context, review strengths and areas for improvement, and iteratively refine their slides before the oral presentation, while also enabling teachers to review automated evaluations, compare them with their own assessments, and monitor student engagement with feedback.

\section{Related Work}

\subsection{Automated Assessment}

Automated assessment encompasses a broad range of computational approaches for evaluating student work. Traditional approaches have long been used in programming education, where automated assessment typically depends on techniques such as test-case execution, static analysis, plagiarism detection, and structured feedback generation. These systems have evolved from simply checking functional correctness to assessing code quality, behavior, readability, security, and teacher-facing learning analytics \cite{paiva2022automated,pieterse2013automated}. In parallel, AI and NLP methods have also been widely applied to text-based assessment, including short-answer grading, essay scoring, and automated feedback through approaches such as semantic similarity, feature engineering, machine learning, and transformer-based models \cite{gao2024automatic}. More recently, large language models have introduced new possibilities for automated assessment, especially in tasks involving open-ended responses. For example, LLMs have been used to generate and administer outcome-aligned assessment activities in informal e-learning settings through instructor-supervised, retrieval-augmented workflows \cite{askarbekuly2024llm}, to support automatic essay scoring and handwritten-text transcription in low-resource contexts \cite{mello2025empowering}, and to synthesize students’ written responses into actionable summaries and misconceptions for teachers through AI-enhanced dashboards \cite{srivastava2025learnlens}.

\subsection{Automated Feedback}

Automated feedback refers to the use of computational systems to provide learners with guidance, hints, corrections, or suggestions intended to improve performance during the learning process. In contrast to automated assessment, which mainly focuses on grading or judging correctness, automated feedback is more directly oriented toward scaffolding learning, supporting revision, and fostering self-regulated learning \cite{cavalcanti2021automatic,deeva2021review}. In particular, automated feedback systems span multiple educational domains and rely on heterogeneous techniques, including rule-based approaches, LLMs, desired-answer comparison, NLP, machine learning, and intelligent tutoring mechanisms. These systems can enhance student performance and improve the speed and consistency of feedback delivery, although challenges remain in demonstrating benefits for instructors, ensuring pedagogical usefulness, and generating feedback that is sufficiently informative rather than merely corrective \cite{hahn2021systematic}.

For example, OpenOPAF illustrates a multimodal approach to feedback on oral presentations that rely on rule-based analysis and template-driven feedback generation, covering aspects such as the use of text in presentation slides and other visual aids, as well as body language, gaze direction, voice volume, articulation speed, and filled pauses \cite{ochoa2024openopaf}. Recent work has increasingly emphasized the role of LLMs in automated feedback. In educational settings, LLM-based approaches have been proposed to generate more adaptive and human-like feedback, but also with the caution that such feedback should be grounded in established learning theories and empirically evaluated rather than relying only on prompt design \cite{stamper2024enhancing}. More broadly, the use of strong LLMs as evaluative agents has also been explored through the \textit{LLM-as-a-judge} paradigm, showing that models such as GPT-4 can produce scalable and explainable judgments on open-ended responses, although issues such as position bias, verbosity bias, and limited reasoning remain important challenges \cite{zheng2023judging}.

\subsection{Learning Analytics Dashboards}

Learning analytics dashboards (LADs) are widely recognized as key instruments for making educational data actionable for teachers, students, and other stakeholders. Dashboards are visual tools that augment human judgment by supporting awareness, reflection, and decision-making through the representation of learner traces and performance indicators. As summarized by \cite{klerkx2017learning}, dashboards should be designed around four core questions: what data are visualized, for whom the visualization is intended, why it is needed, and how the data should be represented and evaluated. Additionally, LADs have evolved from early dashboards based mainly on log data and predictive indicators toward richer approaches that incorporate multimodal data \cite{becerra2025review}, participatory design, and stronger pedagogical grounding, while still facing important challenges related to usability, evaluation, privacy, and adoption \cite{verbert2020learning}. More recently, \cite{topali2025designing} show that human-centered learning analytics and AI in education have strengthened stakeholder involvement in the design of such systems. For instance, recent work has explored teacher-centered and context-sensitive dashboard design through co-design and iterative refinement in real school settings \cite{mohseni2025co, possaghi2025integrating}, the integration of human-centered visual attention analytics based on eye-tracking and AI in online learning environments \cite{navarro2024vaad, becerra2025integrating, sharma2020eye}, and multimodal dashboard systems for MOOCs that combine biometric, behavioral, and log data to support a more holistic interpretation of learner activity \cite{becerra2023m2lads}.

\section{System Description}
\begin{figure}[htbp]
  \centering
  \includegraphics[width=0.8\textwidth]{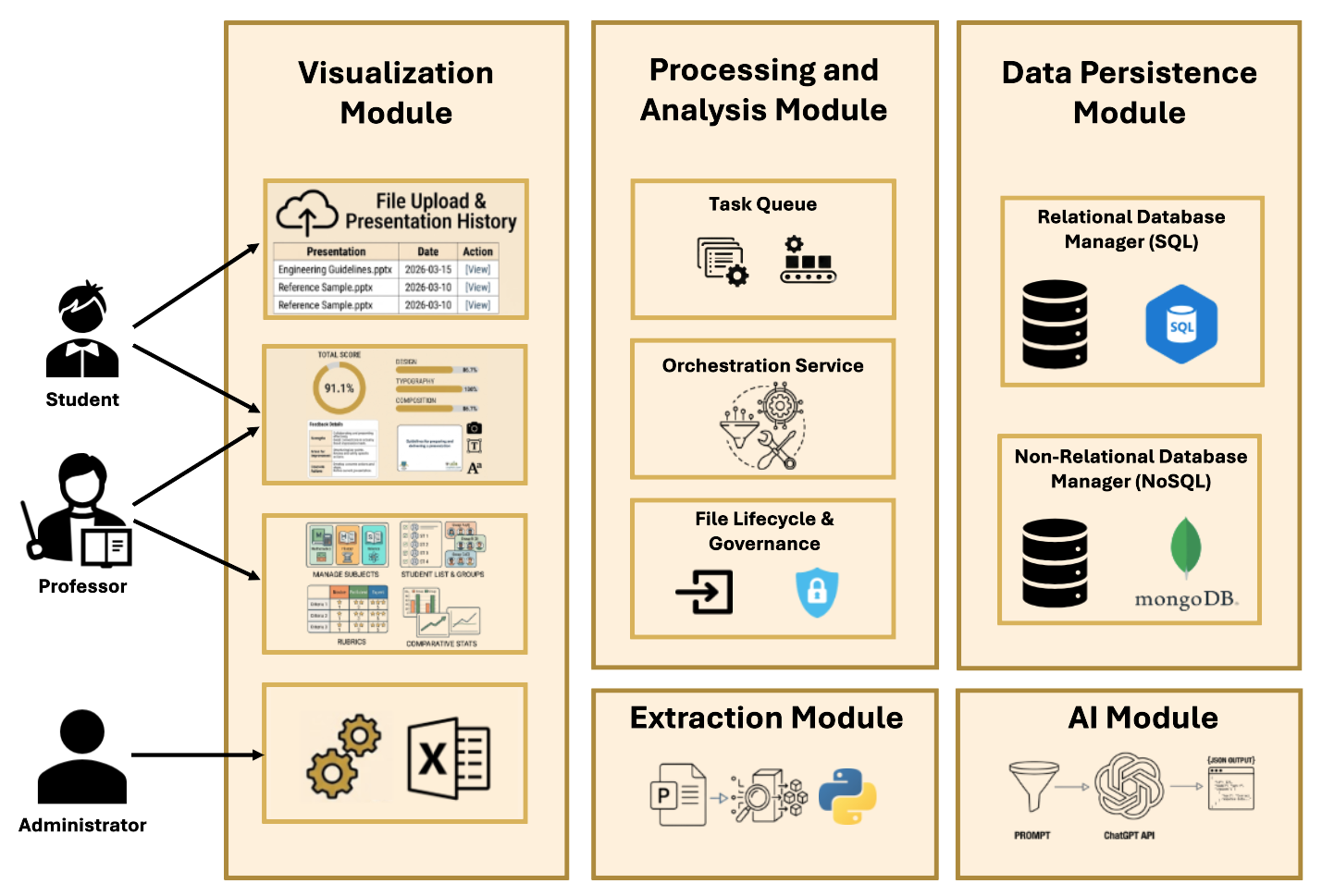}
  \caption{Modular architecture of the AISSA tool.}
  \label{fig:architecture}
\end{figure}

AISSA is built upon a modular architecture developed primarily in Python using Plotly Dash framework. The tool comprises five distinct modules: (1) Visualization Module, (2) Processing and Analysis Module, (3) Data Persistence Module, (4) Extraction Module, and (5) AI Module (Figure \ref{fig:architecture}).

\subsection{Visualization Module}
The frontend of AISSA consists of multiple learning analytics dashboards that provide support for teachers, students, and administrators:
\begin{itemize}
\item \textbf{Student Dashboard:} Allows students to upload PowerPoint presentations (.pptx) or review previous submissions (Figure \ref{fig:student_dash}). From there, they can access a dedicated feedback page within the dashboard, where they are provided with quantitative scores and qualitative feedback. The quantitative score is derived from the AI-based evaluation of each rubric item on a 1--5 Likert scale, using a rubric designed by the teachers. The qualitative feedback is structured into two parts: first, general feedback organized into three paragraphs describing strengths, areas for improvement, and concrete actions to enhance the slides; and second, item-by-item feedback explaining how the student can improve each criterion in the rubric. In addition, the dashboard displays the uploaded slides together with relevant presentation features, such as whether the slides are numbered and the font size used in the text (Figure \ref{fig:analysis_dash}).

    \begin{figure}[htbp]
        \centering
        \includegraphics[width=0.7\textwidth]{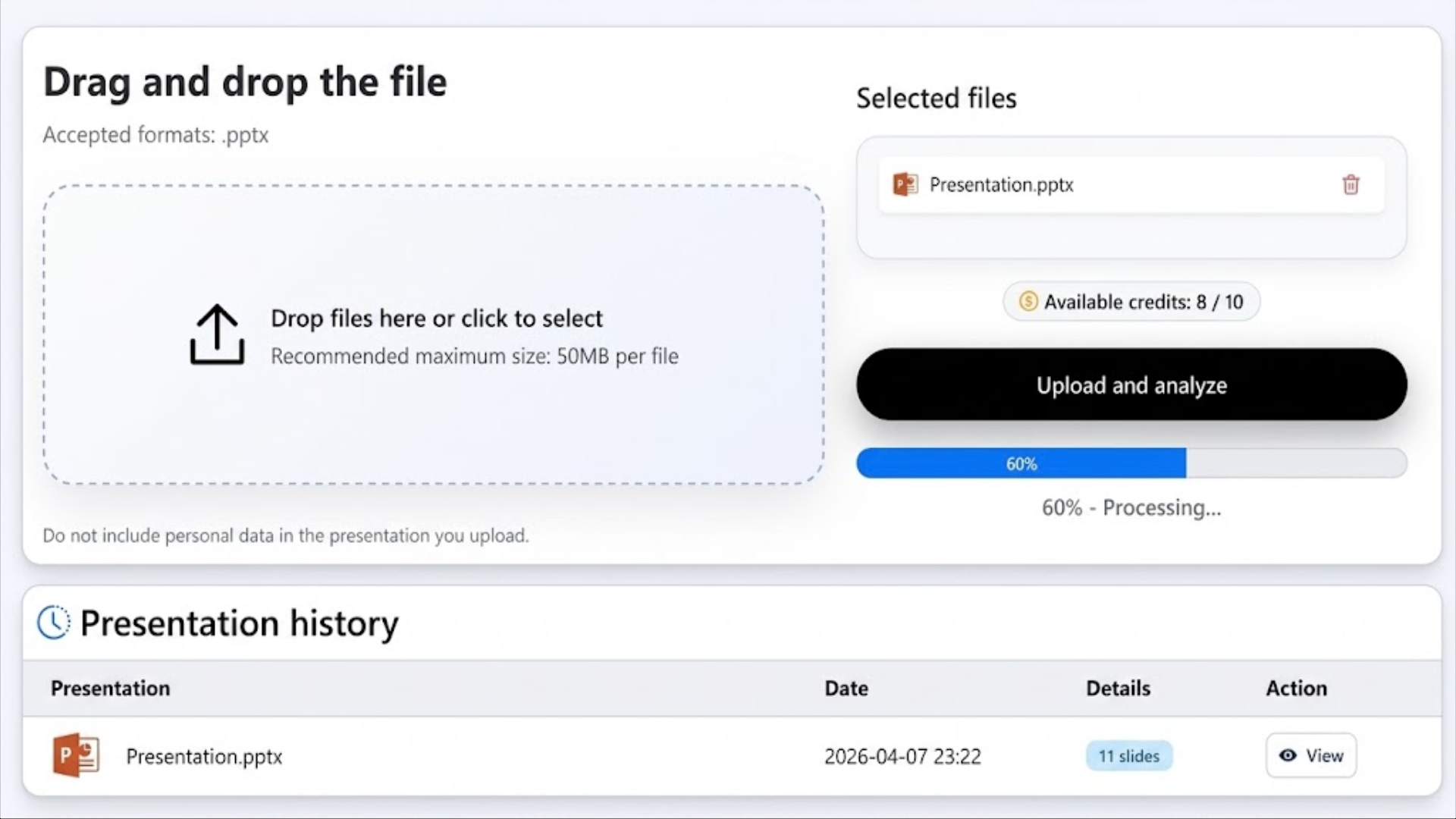}
        \caption{Student Dashboard interface.}
        \label{fig:student_dash}
    \end{figure}

    \begin{figure}[htbp]
        \centering
        \includegraphics[width=0.9\textwidth]{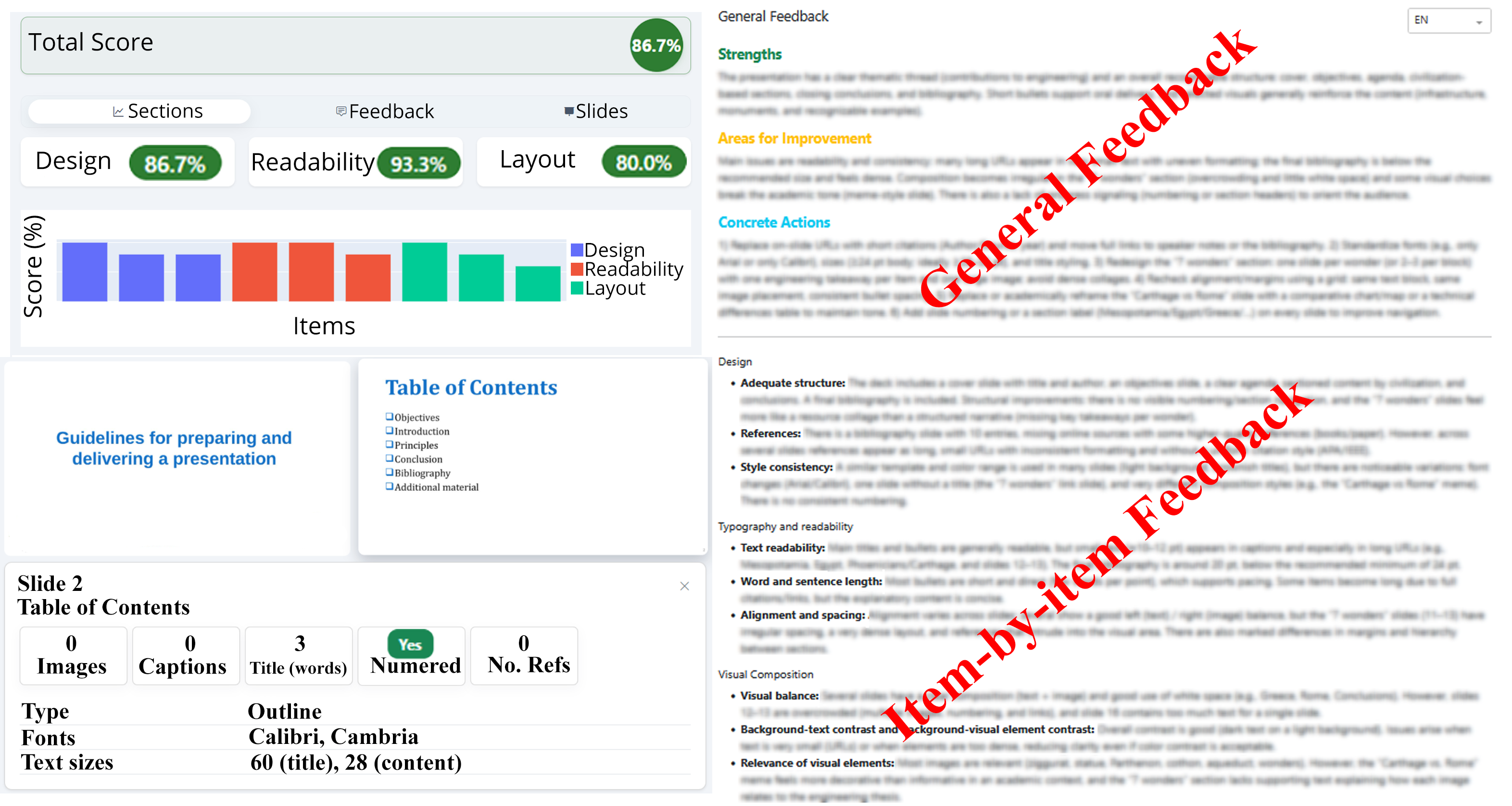}
        \caption{Feedback, Analysis and Slides Visualizations extracted from the Student Dashboard.}
        \label{fig:analysis_dash}
    \end{figure}

\item \textbf{Teacher Dashboard:} Serves as a centralized control interface where educators can configure dynamic 5-point Likert scale rubrics, manage student cohorts, and monitor engagement through activity logs that include indicators such as the number of logins, the number of slide decks uploaded, and the frequency and duration of students’ history review sessions. These logs also enable comparisons between individual student activity and class averages (Figure \ref{fig:teacher_dash}).
    
    \begin{figure}[htbp]
        \centering
        \includegraphics[width=0.85\textwidth]{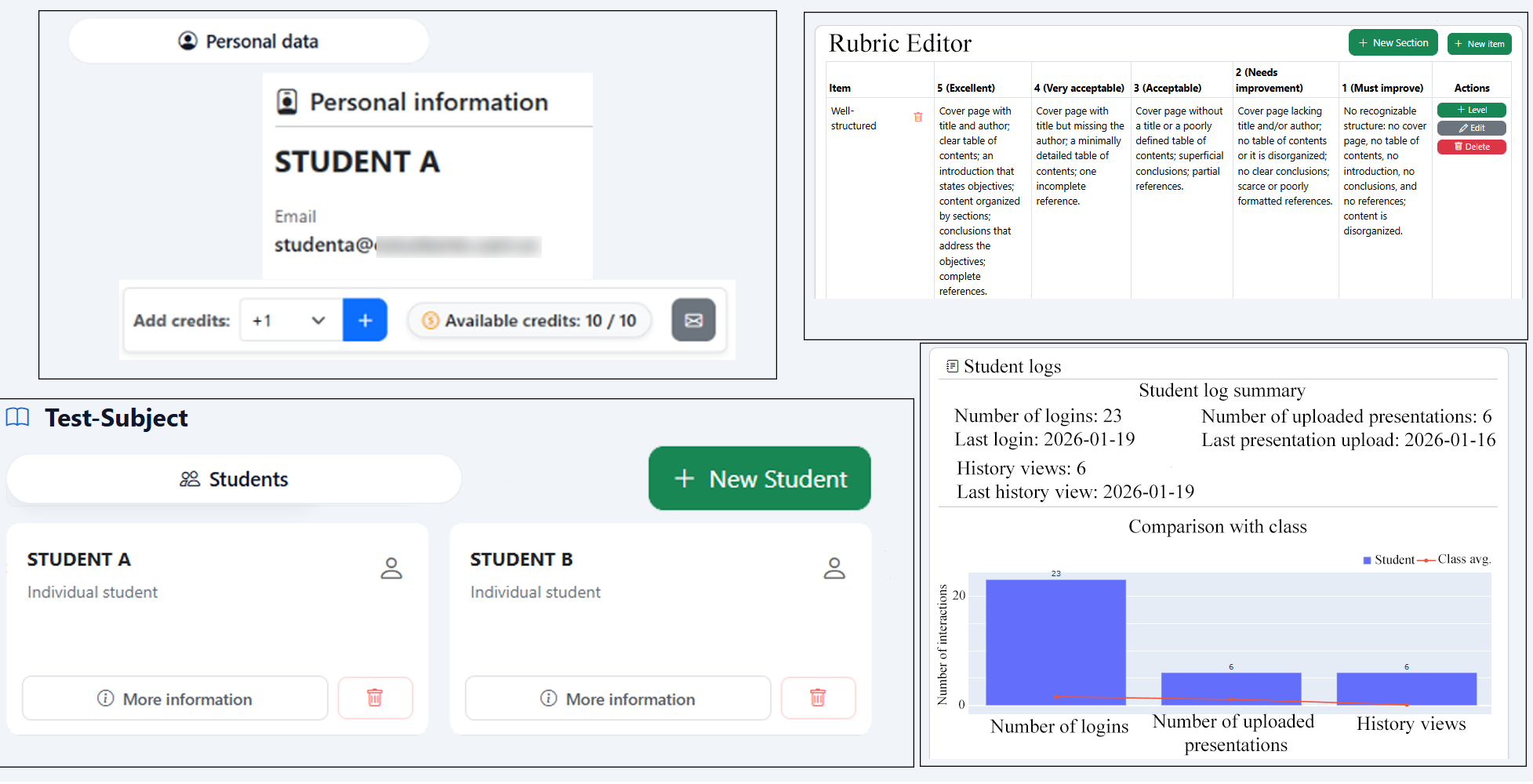}
        \caption{Teacher Dashboard interface.}
        \label{fig:teacher_dash}
    \end{figure}

    \item \textbf{Administration Dashboard:} Centralizes the deployment of the academic infrastructure, facilitating batch imports of student lists, courses, and baseline rubrics via Excel files. 

\end{itemize}

\FloatBarrier

\subsection{Extraction Module}
The Extraction Module is responsible for extracting slide-level syntactic and visual features from uploaded \texttt{.pptx} presentations prior to downstream AI-based analysis. Using \texttt{python-pptx} library, the module traverses the underlying XML structure of each presentation to identify and extract features from each individual slide, including word counts, font sizes, fonts, and reference-related features, such as the number of references and indicators of their quality or type (e.g., hyperlinks, journal articles, books, or legal documents). In parallel, the OpenCV library is employed to analyze graphical content. This analysis includes the characterization of images according to properties such as edge density and color dimensionality, enabling the differentiation of elements such as photographs, logos, and clip art.

\subsection{Artificial Intelligence Module}
\begin{figure}[htbp]
  \centering
  \includegraphics[width=0.8\textwidth]{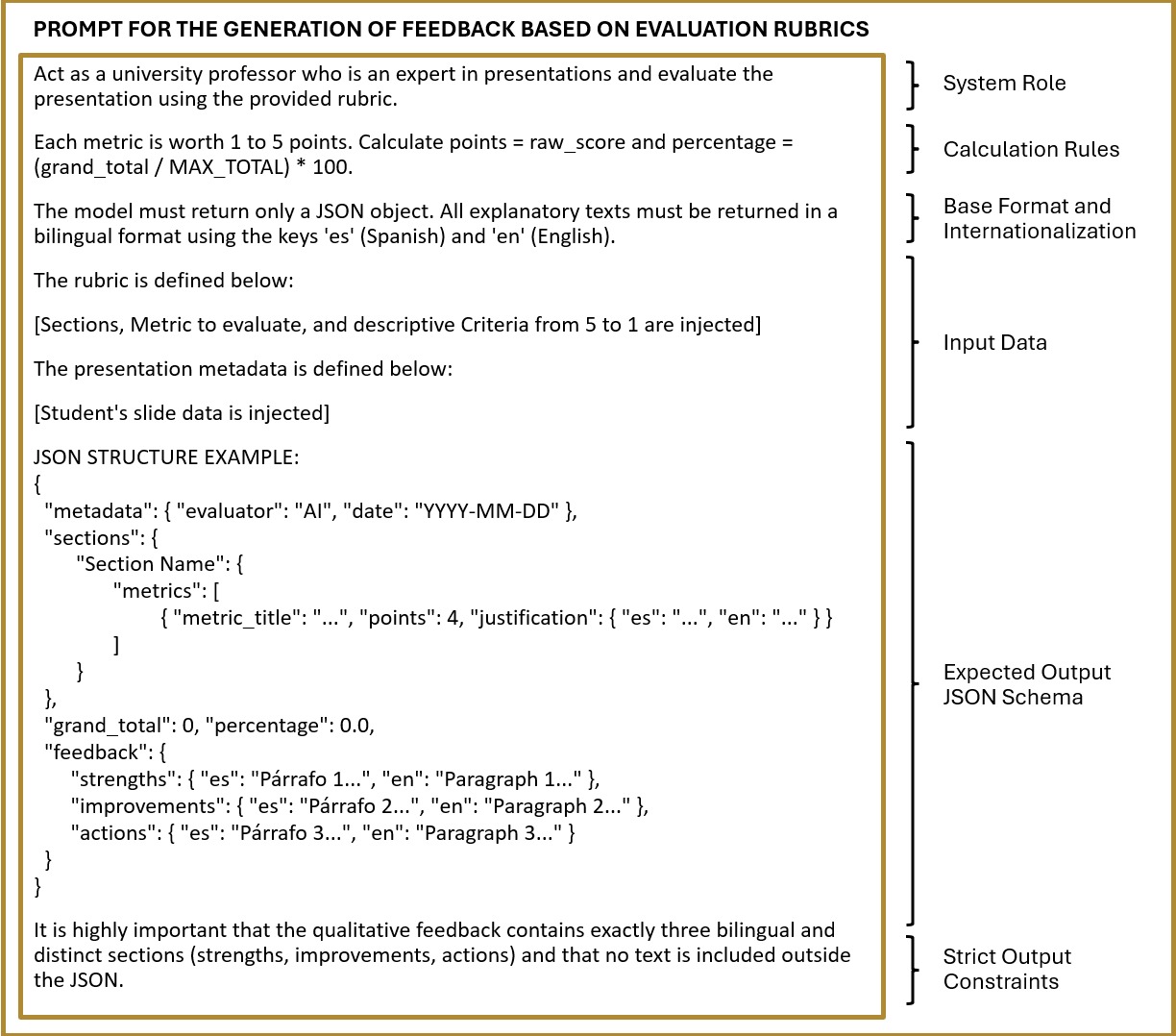}
  \caption{Prompt structure and organization for the AI Module.}
  \label{fig:prompt_structure}
\end{figure}

The Artificial Intelligence Module manages the interaction with the Large Language Model (ChatGPT 5.2) through the OpenAI API. Its main function is to transform the rubric defined by the teacher and the features extracted from the uploaded presentation into structured numerical and qualitative feedback. To this end, the module dynamically builds a restrictive prompt using an adapted version of the GePeTo framework \cite{becerra2024generative}, which combines evaluation instructions, scoring rules, rubric criteria, and presentation-specific metadata (Figure \ref{fig:prompt_structure}).

To ensure a consistent and machine-readable response, the prompt is designed to constrain the model to return a predefined JSON structure. This structure contains, on the one hand, the numerical evaluation of each rubric item on a 1--5 Likert scale, which is then used to calculate the overall score and percentage. On the other hand, it includes bilingual qualitative feedback in Spanish and English. This feedback is organized into two complementary levels: a general assessment divided into three paragraphs (highlighting strengths, identifying areas for improvement, and proposing concrete actions for slide enhancement), following well-established pedagogical feedback structures \cite{hattie2007power, becerra2025enhancing}; and a criterion-level explanation describing how the student can improve each specific rubric item.

As shown in Figure \ref{fig:prompt_structure}, the prompt is organized into several functional blocks. The \textit{System Role} frames the model as an expert university evaluator in presentation assessment. The \textit{Calculation Rules} define the scoring logic used to compute item scores and aggregate results. The \textit{Input Data} incorporates both the teacher evaluation rubric, which specifies the meaning of each level in the 1--5 Likert scale, and the slide-level features extracted by the system. The \textit{Expected Output JSON Schema} specifies the exact structure required by the backend for further processing and visualization. Finally, the \textit{Strict Output Constraints} enforce the generation of valid JSON only, preventing additional conversational text.

\subsection{Processing and Analysis Module}
The Processing Module constitutes the operational backbone of AISSA, orchestrating the execution workflow that connects the visualization, extraction, artificial intelligence, and data persistence modules. After a presentation is submitted, this module receives the uploaded file, stores it temporarily, and coordinates the sequential execution of the extraction and AI services. Once processing is completed, the resulting data are persisted in the database. The module also manages the progress of the evaluation workflow, sending status updates to the frontend so that students can monitor the state of their submission in real time through the dashboard.

To support periods of high demand, particularly during peak submission times, the module relies on an asynchronous processing architecture based on task queues and background workers. Under this design, the web server handles incoming HTTP requests in a non-blocking manner, while computationally intensive tasks, such as slide feature extraction and LLM-based feedback generation, are delegated to background processes. This separation improves system responsiveness and scalability.

In addition, the module supervises key operational services related to authentication, session handling, and role-based access control (RBAC). To prevent redundant executions, it implements SHA-256-based locking mechanisms that avoid the simultaneous processing of identical files. Standard defensive programming practices are also implemented to mitigate Denial-of-Service (DoS) and SQL injection vulnerabilities.

\subsection{Data Persistence Module}
AISSA employs a hybrid database architecture to optimize data storage and retrieval capabilities: PostgreSQL (Relational) manages structured, transactional data, including user credentials, RBAC schemas, course catalogs, the definitions of evaluation rubrics, etc. MongoDB (NoSQL) handles dynamic and semi-structured data. It archives the complex JSON payloads returned by the LLM, the features extracted for each slide, logs, and the physical slide files via GridFS.

\section{Implementation and Technical Validation}
AISSA was deployed in a pilot study at Universidad Autónoma de Madrid (UAM) during the second trimester 2026 within the MOSAIC-F framework \cite{ref-mosaic-f,becerra2026multimodal}. The pilot involved 46 final-year undergraduate students enrolled in the Degree in Telecommunication Engineering. This deployment enabled the assessment of both the technical performance of the platform in a real educational setting and the practical integration of Generative AI for formative feedback on presentation slides.


 \textbf{Processing Efficiency and Cost.} The asynchronous architecture provided stable performance during periods of concentrated use, particularly in the days prior to oral presentation deadlines. AISSA successfully processed a total of 90 presentations over the course without observable service degradation. The end-to-end processing time ranged from approximately 1 to 3 minutes per submission. Regarding the reliability of output generation, no processing errors or data extraction failures were detected, with the model consistently returning the exact required JSON structure. The integration of the LLM (GPT 5.2) through the API also proved economically feasible, with an estimated average cost of \$0.06 to \$0.07 USD per evaluation. This estimate was based on an asymmetric token distribution, with an average of 18,000 input tokens associated with the presentation PDF, rubric definitions, and formatting rules, and between 2,500 and 2,600 output tokens corresponding to the detailed bilingual JSON response.

\textbf{User Experience and Student Use.} To quantitatively assess students' perceived usability of AISSA, the System Usability Scale (SUS) \cite{grier2013sus} was administered to all 46 students, of whom 30 submitted responses. The results yielded an average SUS score of $83.38$, indicating excellent perceived usability. To complement the quantitative findings, semi-structured interviews were conducted with a purposefully sampled subset of 20 students. To ensure a realistic class representation, the sampling sought gender parity and included profiles with varying academic performance and interaction levels. Using personalized inquiries based on activity logs, the interviews revealed that participants found the platform highly intuitive, adopting an iterative workflow of uploading drafts, refining work, and resubmitting without requiring external technical support. Crucially, students perceived the AI-generated feedback as fair and highly useful, particularly praising its precision in improving technical design elements such as font size, visual consistency, and bibliography formatting. Nevertheless, several students reported intentionally disregarding some recommendations when they considered that those suggestions conflicted with their planned visual design or presentation strategy.

\section{Conclusion and Future Work}
This paper presented AISSA (AI-based Student Slides Analysis tool), a web-based tool that combines LLMs and Learning Analytics dashboards to provide scalable, rubric-based feedback on oral presentation slides. By integrating automated scoring, structured qualitative feedback, and interactive visualizations for both students and teachers, AISSA addresses key challenges related to feedback quality, timeliness, and scalability in higher education.

The pilot suggests that AISSA is both feasible and useful in practice. At the technical level, the system operated reliably in a real academic deployment, with stable processing and consistent structured output. At the educational level, the first evidence indicates that students used AISSA iteratively: they uploaded slides, reviewed the generated feedback, revised their materials, and resubmitted improved versions. Students especially valued the feedback on formal and design-related aspects.

Overall, AISSA shows the potential of combining LLM-based analysis with Learning Analytics dashboards to support scalable formative feedback on presentation slides before the oral presentation takes place.

Future work will focus on four main directions. First, we will compare AISSA’s LLM-based evaluations with assessments provided by human teachers, examining both the numerical scores assigned at rubric-item and overall level and the qualitative feedback associated with each presentation. This comparison will make it possible to study agreement not only in grading outcomes, but also in the content, specificity, and pedagogical value of the generated feedback. Second, we will study the pedagogical impact of the tool more systematically, including whether iterative use of AISSA leads to measurable improvements in slide quality and final oral presentation performance. Third, we will continue refining the AI pipeline by comparing different LLMs, including both proprietary and open-source models, and by extending the set of extracted slide features. Fourth, we will further examine students' selective uptake of AI-generated recommendations and we will analyze whether students who selectively apply AISSA feedback achieve better outcomes than those who follow all recommendations or disregard them entirely.

\section*{Acknowledgements}
    Support by projects: Cátedra ENIA UAM-VERIDAS en IA Responsable (NextGenerationEU PRTR TSI-100927-2023-2), M2RAI (PID2024-160053OB-I00, MICIU/FEDER) and SNOLA (RED2022-134284-T). Alvaro Becerra is funded by a predoctoral contract (FPI) from the Comunidad de Madrid (PIPF-2024/COM-34288).
\bibliography{sample-ceur}

\end{document}